\documentclass{article}

\usepackage[english]{babel}
\usepackage{makecell} 
\usepackage[letterpaper,top=2cm,bottom=2cm,left=3cm,right=3cm,marginparwidth=1.75cm]{geometry}

\usepackage{amsmath}
\usepackage{graphicx}
\usepackage[colorlinks=true, allcolors=blue]{hyperref}
\usepackage{lmodern}
\usepackage{ragged2e} 
\usepackage[square,numbers]{natbib}

\title{SCOPE: Simple Coil Optimization for Plasma and Engineering}
\author{N. Welch and C. Marsden}

\begin{document}
\maketitle

\begin{abstract}
Designing superconducting coils for a tokamak fusion device is a highly coupled, non-linear design problem. The coils have many disparate engineering requirements from structural to power electronics, as well strict limits placed on the system by the high temperature superconducting (HTS) cables. Simultaneously, the coils must be able to contain multiple plasma scenarios from inception, through ramp up, to flat top, and ramp down, all whilst applying a large, controlled, inductive voltage to drive current. In addition, we wish to optimize divertor separatrices to increase the likelihood of designing a suitable divertor strikepoint. Lastly, the physical limits of the entire tokamak must be taken into account and space reserved for support structures, access for maintenance schemes, and installation limits.

The method outlined here uses a combined simulated annealing method to find optimal coil sizes and positions with a constrained quadratic or quartic optimization for the coil currents. The method is designed to optimize coils for multiple scenarios simultaneously, including ramp-ups, to avoid over optimization of a single design point. A key enabler is the efficient implementation that allows millions of evaluations to be performed in a few hours with modest computational power.

This optimization method is part of a larger, iterative workflow which enables further, detailed design work to feedback on the optimization.
\end{abstract}

\section{Introduction}

The interaction between the magnetic coils and the plasma must be accurately simulated and understood in order to ensure the successful achievement of the tokamak mission. This requires tremendous design activity to ensure the plasma has the correct performance criteria as shown in the FPP pre-concept design paper, 
\cite{STE1_overview_paper_PLACEHOLDER, STE1_flat_top_paper_PLACEHOLDERR, STE1_ramp_up_PLACEHOLDER}, often using a mixture of fixed and free boundary equilibria \cite{Metis_overview_2018, FreeGS_documentation_2025, Metis_FEEQS_coupling_2021}.
While these methods allow for successful plasma modelling, it has no guarantee of allowing realisable magnets. Due to the very high fields and current densities, the only practical solution for the three primary magnet systems: the toroidal field (TF) coils, the poloidal field (PF) coils and the central solenoid (CS) is high temperature superconductors (HTS) as seen in other tokamak designs \cite{M_Windridge_TE_and_HTS_spherical_tokamaks_overview_2019, STEP_special_issue_TheMagneticCage, ARC_overview_2015, HH70_overview_2025, sparc_overview_2020, sparc_TF_program_2024}. Whilst HTS is an enabling technology, it comes with its own strict limitations and engineering concerns, chiefly the field limited critical current, \cite{SuperconComparison}, as well as other concerns discussed below.
Many attempts have previously been made to optimize the PF and CS coils for a given equilibrium whilst also optimizing the equilibrium \cite{EAST_PF_optimisation_genetic_algorithms_2006, PF_optimisation_using_BLUEPRINT_2020, Nilima2024_FusionEngDes_STEP_Bluemira_PFcoil_Optimisation, bayesian_optimisation_of_PF_coils_STEP_2025, Bardsley2024_PPCF_Decoupled_Magnetic_Control_ST_Divertors}. These tended to have involved limited optimization for the coils, only focussing on one of the simplest engineering criteria and typically only focussed on a single equilibrium whilst omitting the ramp up scenario and the inductive flux swing which is both critical to the scenario and also a dominating feature of engineering design.

\section{Structure of the paper}
We begin with the motivation for this optimization tool from the combined plasma-magnets-divertor workflow and how these linked areas must work both consistently and also independently to avoid being waylaid by issues in any of these disparate areas. This will include an outline of what SCOPE is and how it helps this workflow. Then we will go into the constraints on the system and the optimization terms used to minimize potential engineering issues. We then discuss in detail the optimization methodology used in SCOPE before remarking on results from the ST-E1 pre-concept design.

\section{Combined Workflow}
The PF, CS, plasma must all be designed to work together and so require a very involved parallel workflow, as outlined in \cite{STE1_integration_paper_PLACEHOLDER} with \ref{fig:} showing the poloidal cross-section of an example ST-E1 pre-concept design point. For the purposes of this paper we will assume that the system is also vertically symmetric, in that the positions of the coils are vertically symmetric between the top 5 PF coils and the bottom 5, and the central solenoid will be similarly symmetric. Note that we also use a segmented solenoid with varying current, however for clarity in this paper we report the maximum loads such as field and stresses. The underlying method outlined here has been developed to work on vertically asymmetric designs and many of the ST-E1 scenarios have been asymmetric in nature. Becuase of this, we allow the currents to be asymmetric, as even for nominally symmetric FBEs there can be slight asymmetries. 

The motivation for creating SCOPE was to allow a single point in this workflow to combine multiple considerations at once. However, this is not to create a monolithic tool to design an entire tokamak, instead it has just enough detail to allow the complex worlds of plasma physics, superconducting magnets and divertor engineering to have their primary design issues addressed simultaneously. We then use SCOPE as a tool that allows iterative design of these three areas so that any area can be updated independent of the others and crucially give useful feedback to all areas. This feedback cannot simply be a single go/no-go but instead considers multiple different scenarios simultaneously to give more directed feedback on what is possible and what is not, what is easier and what is more difficult, including crucial quantities such as possible inductive flux swing. In this way, an iterative design methodology is much more likely to converge on a feasible design point. To allow this, SCOPE was designed to avoid over-optimization for a single  scenario. Instead, we optimize for multiple different flat top scenarios simultaneously, keeping the same coil geometry and finding different coil currents. Similarly, SCOPE can optimize for multiple static equilibria that create a pseudo-time domain plasma ramp up and flat top. 

In addition, the performance and loads on the PF and CS coils are often highly coupled. The PF coils at lower radii are physically close to the top of the solenoid and the magnetic fields created by the two are likely to be very similar. This will result in coils which will create large fields on one another, which both limits HTS critical currents but also increases stresses. Additionally, they will be highly inductively coupled limiting quench performance as they increase the stored magnetic energy and require higher voltages to drop their currents quickly.

Therefore, due to the closely coupled nature of the PF and CS coils, we treat them equivalently. This also allows us to include the inductive flux swing in the optimization, rather than optimizing for the plasma confinement and divertor geometry and assuming a flux swing will have little effect on these. Instead, we can take into account the full combined inductive and field coupling of all the coils, which is crucial in order to minimize the total magnetic field that will reduce stresses and increase critical currents in superconductors.

Given this, we still only treat each coil as a simple, single turn object with a total current. We do not directly calculate the superconducting critical currents as this requires detailed cable-level study to first find copper fractions or space for cryogenics or internal and external supports. Instead, by reducing related quantities such as magnetic fields and stresses, we greatly increase the likelihood of a detailed engineering design being achievable. If it is not, then SCOPE has many optimization weights that allow the magnet design to feedback and alter in order to reduce the biggest issues, again speeding up the number of iterations required for design.

Secondary systems can also be coupled with this workflow using outputs from SCOPE such as currents, voltages and inductances which aid in early designs for the cryogenics and power systems, including required electrical power from power supply units (PSUs) and potential current lead designs.

SCOPE uses an 'inner' constrained analytic cost-function minimization for the coil currents, where the cost-function is used by an 'outer' simulated annealing algorithm which optimizes the coil sizes and positions. This method was chosen to allow fast optimization of the continuously valued currents and cost-function, coupled with global position optimization that is constrained by literal and figurative hard walls.

\section{Plasma Constraints}
The chief requirement on the coils is to shape the plasma. Exactly what is required from the magnets to do this will be provided by the free-boundary equilibria (FBE) which in a tokamak is a description of the poloidal flux required from the coils. Using cylindrical symmetry, this is given by:

\begin{equation}
\psi(r,z) = 2\pi\int_0^r r'B_z(r',z) dr'
\end{equation}

where $B_z$ is the vertical component of the flux density, integrated over a disk centred on the machine axis, $r=0$, extending to some radius $r$. In this paper we will be explicit regarding the source of the flux, as the total is given by the CS, PFs, and plasma current.

In order to greatly decrease the computational workload in SCOPE, we assume that if the poloidal magnetic flux distribution around the last closed flux surface (LCFS) created by the coils in the FBE is accurately recreated by any set of coils and currents then the flux distribution and the plasma will similarly be accurately recreated inside the LCFS. This is based on the inability for coils to create non-local field maxima or create any field variation with faster spatial variation than the LCFS-coil distance \cite{ErnshawTheorem}. Therefore, we can place a number of constraint points around the LCFS, much like in a typical FBE simulation, and ensure that the flux at these points is recreated to within some tight tolerance. To validate this assumption, we can re-run any FBE with the optimized coils and currents, this also allows us to set the tolerance required, using the following constraint function:

\begin{equation}
    c(t) = \sum_i w^c_i [\psi_i^{FBE}(t) - \sum_jG_{ij}I_j(t) + \Delta\psi(t)]^2
    \label{eq:fluxConstraint}
\end{equation}

where $w^c_i$ is the weight around the LCFS at position $i$, $\psi_i^{FBE}$ is the flux from the FBE, $G_{ij}$ is the Green's function for the flux from coil $j$, $I_j$ is the current in coil $j$ and $\Delta\psi$ is a spatially constant shift found later. Note that these can all be time dependent quantities for a set of equilibria modelling a time evolution of a plasma or a set of unrelated equilibria to allow a study of multiple simultaneous flat-top scenarios, as described above.

\begin{figure}
\centering
\includegraphics[width=0.5\linewidth]{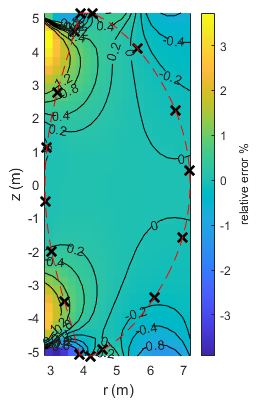}
\caption{\label{fig:fluxRecErr} Surface plot of the typical percentage difference between the flux created by the coils from the FBE and the flux after optimization. The LCFS (red dashed) is shown with the constraint points (black crosses). At maximum the deviation of the flux recreation reaches 1.4$\%$ with 0.25$\%$ set as the average by the constraint function.}
\end{figure}

Further, we define the constraint functions using:
\begin{equation}
    C(t) = c_\textrm{tol} - c(t)
\end{equation}

such that the constrained minimization, described later, will be met when $C \geq 0$ and won't allow $C<0$. From previous trials, it was found 0.25\% allowed good recreation of the original FBE, ensuring that key features such as the X point placement and exact in-board and out-board midplane radii. Figure \ref{fig:fluxRecErr} shows the typical error distribution when applying this limit, while certain points around the LCFS may have larger error, reaching a maximum of just over 1$\%$, the majority is less than 0.2$\%$.

Note, our coupling of these constraints and coil optimization, is not equivalent to the regularization used in some FBE solvers, as they typically attempt to reduce either the L2 norm of the current updates between convergence steps or the total current in the coils. As this is applied during a non-linear convergence, there is no guarantee that the currents produced at convergence will be reduced. What is more likely is that the convergence will require many more steps to converge.

Previously, there has been a method that uses a spherical harmonic basis rather than specific points in space, \cite{Bardsley2024_PPCF_Decoupled_Magnetic_Control_ST_Divertors}. This approach was not used as it was found to be very difficult to get sufficiently high accuracy with a limited number of harmonics, especially for the X point . In particular this is difficult to accurately recreate as it is a very sharp spatial feature which requires high order harmonics and also an example of catastrophic cancellation where sizeable field contributions from the individual coils and the plasma must all cancel out to zero. Therefore, small relative errors in each contribution can lead to large errors in the sum and a shifting of the X point. This then breaks the assumption that the plasma current will retain the same distribution.

\subsection{Inductive Drive}
In addition to confinement via recreating the FBE LCFS, we make the assumption that we can add or subtract a spatially constant flux from the entire LCFS, $\Delta\psi(t)$ in equation \ref{eq:fluxConstraint} above,  without disturbing the plasma current at all. This is due to the normalized manner in which FBE parameterize the plasma. However, the variation of the LCFS flux value in time is crucial to creating the required inductive voltage in the plasma. Typically, a set of quasi-static FBE are created using information from a time-domain fixed boundary equilibrium. In this fixed boundary evolution the various current and heating sources are described and some internal plasma transport is taken into account. These time-domain simulations will produce a desired inductive voltage at different points in time. We can then include these in the optimization first by defining the inductive voltage felt by the plasma: 

\begin{equation}
    V_{\textrm{ind}}(t) = -\frac{d\psi_{\textrm{LCFS}}}{dt}
\end{equation}

and then by integrating to give the required temporal behaviour of the LCFS:

\begin{equation}
    \psi_{\textrm{ind}}(t) = -\int_0^t V_{\textrm{ind}}(t') dt' + \psi_0
    \label{eq:fluxtime}
\end{equation}

where we are free to add a constant of integration, in this case a flux value, $\psi_0$, that is constant spatially and in time. Recreating this time varying flux and choosing $\psi_0$ will be critical in optimizing the coils as the inductive drive will be a primary cause of loads on the PF and CS coils and a key area for feedback between the plasma and magnet teams.

\begin{figure}
\centering
\includegraphics[width=0.5\linewidth]{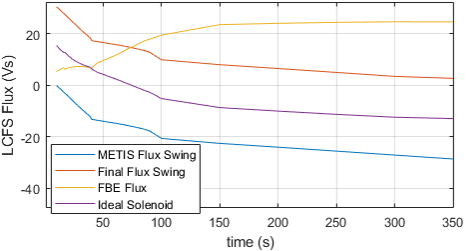}
\caption{\label{fig:fluxVals}Showing different flux on the LCFS for the optimization ramp up case (blue) the flux swing from METIS required for the 27.8 Vs total flux swing in time assuming no constant of integration, (red) the flux on the LCFS after optimization including a flux shift, (yellow) the value of flux on LCFS from the bare FBE that had to be cancelled and (purple) the flux from the solenoid using the infinite solenoid approximation by integrating the total flux through the solenoid bore at mid-plane.}
\end{figure}

As the FBE simulations treat each equilibrium as a single snapshot in time, often the value of flux on the LCFS has no real meaning and the FBE solver just converges when a flux contour sits on the prescribed spatial points. When creating the pseudo-time ramp from these, a smooth variation of LCFS flux was seen, yellow line in figure \ref{fig:fluxVals}. This smooth variation was largely due to using the previous FBE as an efficient start point for the next FBE in the pseudo-time series. However, this positive variation in flux, driven by the increase in plasma current and positive field from the coils, if accurately recreated, would drive a reverse voltage, reducing the flux swing and ultimately decreasing the plasma current. Therefore, care must be taken to cancel this flux variation and ensure that the flux follows the time variation given in eq. \ref{eq:fluxtime}.

From these we can then set a time-varying offset to the constraint given above:

\begin{equation}
    \Delta\psi(t) = \alpha\psi_{\textrm{ind}}(t) - \psi_0 -\psi_\textrm{FBE,LCFS}(t)
\end{equation}

where $\alpha$ is a factor that we manually set for each optimization to help ascertain what flux voltage will be possible with a set of coils, ie we will optimize with $\alpha = 1$, then if the coil engineering shows that a ramp is not possible we can reduce this to some value until a feasible coil set is produced. We can then inform the plasma workflow of what flux swing is possible. In tandem with this, we will also manually vary $\psi_0$ to reduce the loads on the coils, typically by getting the fields on coil to be equally large at the start and end of a ramp up. The effect of this is seen in figure \ref{fig:fluxVals} where the final flux swing (red) is shift to start positively from around 40Vs and then drop to around 0.

When optimizing a set of flat-top equilibria in psuedo-time, we don't have any of this information, so we either set $\psi_0 = -\min[\alpha\psi_{\textrm{ind}}(t)]/2$ or we set $\psi_\text{LCFS} = 0$, the latter usually produces a more symmetric swing for the solenoid as discussed in the results section.

\section{Spatial Constraints}
We do not wish to limit the coils to pre-determined regions in space as this will likely result in local optimization where the initial results are only slightly improved without reviewing all possible coil placements.

\begin{figure}
\centering
\includegraphics[width=0.5\linewidth]{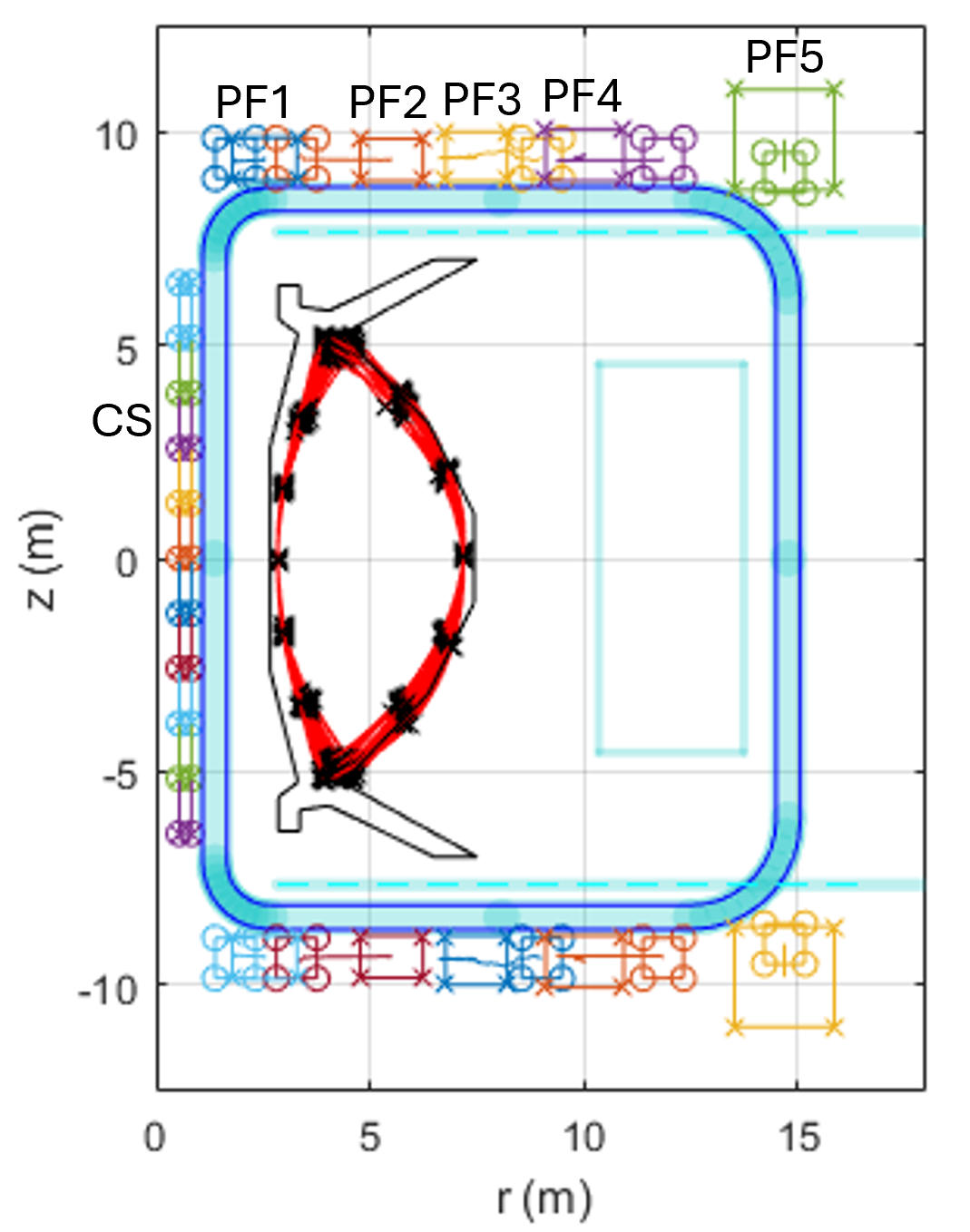}
\caption{\label{fig:} Axisymmetric  of an ST-E1 pre-concept design point. A set of representative plasma LCFS (red lines) with constraint points (black crosses). The TF coils (dark blue) are the primary structural object that the coils must be designed around. The path taken by the initial PF coils (circle-lines) to the optimized coils (cross-lines) is shown. The light blue regions show the spatial constraint areas and can be added arbitrarily. Note the internal structures (black lines) are inconsistent with the plasma due to the parallel development of multiple simultaneous workflows.}
\end{figure}

This being said, the coil optimization must take into account physical objects that the coils cannot overlap with. Chiefly, the TF coil which is described by an outer boundary and a user dictated separation distance required for the TF structure. However, we can take into account any other spatial object or limit by using constraint lines and a minimum approach distance. To do this, during the simulated annealing step, described below, we check all coil-line closest separation and if a coil is within a minimum distance then we abandon the step and start another. Additionally, if the initial coils in an optimization break these constraints, then we continue the step but add a large penalty term to the cost-function, described below, which is proportional to the inverse of the line-coil separation. This then makes it exponentially more likely for the coil to move outside of the constrained areas. Once all constraints are satisfied, the previous method is applied so no coil can re-enter a constrained area. This allows iterations of new tokamak geometries to be used with previous versions of coils and plasmas. Additionally, a check is made to ensure that coils do not overlap with each other. If they do then the change is abandoned and another trial begins.

\section{Coil Engineering}
The previous section described hard constraints that are applied to the system. In order to gauge the performance of the coils, we create a cost-function which we will attempt to minimize but will not strictly limit the loads on the coils. This approach is taken as the performances of the coils are very interrelated, so simple limits that treat the coils in isolation such as maximum currents aren't useful as they don't take into account how neighbouring coils will effect each others field sensitive HTS critical current. In addition, there are few very set rules for coil engineering and many are not easily stated without designing the coils further. These terms are also deliberately created to be analytic in terms of the coil currents so that a fast optimization method can be used to find optimum solutions. Therefore we aim for 'monotonic proxies' which are efficiently calculated quantities that are simply related to the more complex engineering issues and by reducing the simple measure we are very likely to reduce the actual engineering issue.

Also, note that most of the optimization terms in the cost-function take the same form of a metric squared  then multiplied by a weight term that is the inverse square of the units of the metric. This allows us to add multiple terms together to create a total cost-function. The squaring-then-summing ensures that we are minimizing the loads for each individual coil and, by squaring the metric we are preferentially reducing the loads on the worse effected coils during the optimization.

\subsection{Shape and Volume}
The simplest engineering considerations are the geometry of the coils. The majority of the following optimization terms will want to make the coils much larger, with only a weak drive to make coils smaller to minimize plasma - coil current centre distance. Therefore, we have to add a term which is independent of the current and calculated directly:

\begin{equation}
    \Phi_{\textrm{vol}} = \sum_i w_i^\textrm{vol}\pi R_i^2 \Delta R_i \Delta Z_i
\end{equation}

where $w_i^\textrm{vol}$ is the inverse of the desired volume for each coil. The coils are given by their centroid vertical position and radii, $Z_i$ and $R_i$, and their heights and widths, $\Delta R_i$ and $\Delta Z_i$.

In addition, we can add hard constraints to any of these values during the simulated annealing step, described below.

\subsection{Current}

The current is the simplest quantity that we can analytically optimize, using the following Tikhonov regularization:

\begin{equation}
    \Phi_{\textrm{current}} = \sum_i w_i^\textrm{current}I_i^2
\end{equation}

where $w_i^\textrm{current}$ is the inverse square of the desired current in each coil. Note that we could easily make this optimization based on current density by adding in the inverse of the area of each coil to this weight factor. As the coils often reached their maximum allowable size, these approaches become identical.

\subsection{Field}\label{Field}

The field on each coil is not directly a limiting factor. However, it works well as a proxy for critical current density, especially as we can take into account the directional dependence of the field:

\begin{equation}
    \Phi_{\textrm{field}} = \sum_i \sum_j\left(w_i^\textrm{Br} \left[G_{ij}^rI_j\right]^2 + w_i^\textrm{Bz}\left[G_{ij}^zI_j\right]^2\right)\Delta_i 
\end{equation}

where $w_i^\textrm{Br}$ is the inverse of the desired radial field component squared, similarly for $w_i^\textrm{Bz}$ and the z component. Here we integrate over the edges of the coil at positions $i$ with integral weights $\Delta_i$, where $\sum_i\Delta_i = 1$. By integrating around the edges, we are likely to get a representative of the maximum field on the coil and avoid any numerical issues with integrating inside each coil. This is then repeated for each coil.

\subsection{Forces and Stresses}

After superconducting considerations, chiefly the critical current, the next biggest issue facing the PF and CS coils are the very large Lorentz forces that are produced from large magnetic fields and current densities.

This can be done in a number of ways, considering either force densities, total forces or pressures. These are all considered by cylindrical integrals of the form:
\begin{equation}
f_i = 2\pi\int_{R_i-\Delta R_i/2}^{R_i+\Delta R_i/2} \int_{Z_i-\Delta Z_i/2}^{Z_i+\Delta Z_i/2} r\underline{g}_i (r,z). (\underline{J}_i \times \underline{B})  \:dr\:dz
\end{equation}
where we're integrating over the volume of coil $i$, with a spatially varying vector function, $\underline{g}_i$ using the current density of the coil $\underline{J}_i$ and the total magnetic field from all coils, $\underline{B}$.

For example, if $\underline{g}_i  = \underline{\hat{z}}$ then we have the integrated vertical component of the force. If we use $\underline{g}_i  = \underline{\hat{r}} \frac{1}{\Delta R \Delta Z}$ then we have a simple approximation of hoop stress.

As previously, we create the cost-function terms by using the square of the terms with some weighting based on the inverse square of the desired behaviour:

\begin{equation}
    \Phi_{\textrm{forces}} = \sum_i w_i^\textrm{forces}f_i^2 
\end{equation}

Typically, we found that hoop stress was the key quantity to minimize for, however vertical forces in the solenoid can also be very problematic.

\subsection{Magnetic Energy and Enthalpy}
As the coils in a fusion relevant tokamak are large and carry sizeable currents, they naturally have immense stored magnetic energies. Dealing with this is an issue for any electromagnetic system (eg Bitter magnets) however for superconducting coils there are additional ramifications due to the coils quenching. This is when the coils, for a range of reasons, cease being superconducting which causes current to preferentially travel through any other connected conductors, typically the large amounts of copper in the system. This causes additional heating and often causes a run away effect, typically referred to as a quench.

A simple metric for a coil is the stored magnetic energy associated with that coil divided by the energy required to heat the coil to a large but not damaging temperature. The idea is that if a quench occurs, either the coil will passively begin to heat the coil or an active system will remove current and magnetic energy from the system. For either system, the ability of the coils to store thermal energy relative to their stored magnetic energy will be critical. This does not aim to model a quench, just to increase the likelihood that optimized coils will be quench protect-able with further engineering design.

The cost-function term for this is:

\begin{equation}
    \Phi_{\textrm{therm}} =  \sum_i\left[ \frac{I_i\sum_j L_{ij}I_j}{H_{\textrm{therm}}2\pi R_i^2 \Delta R_i \Delta Z_i}\right]^2 
\end{equation}

where $L_{ij}$ is the mutual inductance between coils $i$ and $j$, we also include the typically larger self inductance in this. For $H_{therm}$, we use the volumetric enthalpy of copper from a nominal operating point of 20K to 120K, along with the coil volume, this then acts as the weighting for this term and the maximum temperature can be altered if necessary. Note that we do not calculate the total energy here and instead the second sum over all coils is carried out after the square, so we are minimizing the magnetic energy 'associated' with each individual coil relative to its thermal mass.

\subsection{Power Supply Voltage}

For a time-domain ramp up, the voltage required to ramp the coils may be prohibitive. Whilst the single-turn approximate design used here to optimize coils does allow the current-voltage characteristics to be varied by using a variable number of turns in each coil, the total turn-turn and lead-lead voltages may still be dangerously high for a reasonable number of turns. For superconducting coils large voltages will likely be down to the inductive voltage required to create or modify the large magnetic fields, primarily when the coil currents change rapidly.

One of the chief reasons for rapid current changes is due to the lack of true time-domain modelling in the FBE and optimization and as such the currents in the coils can vary wildly in time. As mentioned above, using a sequential method for the FBE creation can greatly reduce this but will not reduce the problem entirely.

To optimize for this, we need to introduce a time differential in the optimization:

\begin{equation}
    \Phi_{\textrm{voltage}} = \sum_i w_i^\textrm{voltage} \left[\sum_j L_{ij} \frac{dI_j}{dt}\right]^2
\end{equation}

where, as usual, the weighting term is the square inverse of a desired voltage. In order to calculate this we use a central difference method for the differential. In order to achieve a global minimum for the currents, this term requires all currents to be optimized for simultaneously, whereas all previous cost-functions allow for currents to be optimized at each point in time separately. 

\subsection{Transient Field Variation}
Much like a superconducting quench, we cannot hope to include AC losses in the HTS in this optimization scheme. However, our philosophy of reducing related quantities does allow us to reduce the variation of magnetic field in time. This is done by minimizing the time-domain variation of magnetic flux density on the coil edges, using much of the same methodology as section \ref{Field} except using a central difference method for the differential of the current:

\begin{equation}
    \Phi_\textrm{dBdt} = \sum_i \sum_j\left( \left[w_i^\textrm{dBrdt}G_{ji}^r \frac{dI_i}{dt}\right]^2 + \left[w_i^\textrm{dBzdt}G_{ji}^z\frac{dI_i}{dt}\right]^2\right)\Delta j 
\end{equation}

where, again, we can use different weights, $w_i^\textrm{dBrdt}$ and $w_i^\textrm{dBzdt}$, to target the different components separately. Similarly, to the inductive voltage optimization, this will require all currents to be optimized in time simultaneously, rather than optimizing each quasi-static equilibrium separately.

\subsection{Divertor}

The flux separatrices forming the divertor legs can also be optimized in a similar manner to the coil performances, in that no hard constraints are set, instead a desired behaviour is stated and the coil currents are optimized to approach this.

Firstly, the separatrix legs can be shaped to follow a given line ending at a desired strike point by stating various points along the path and ensuring that the total flux is equal to the optimized LCFS flux value, $\psi_\textrm{LCFS}^\textrm{opt}$, taking into account the flux swing and shift as described above. To do this, we require the flux from the plasma current calculated at these chosen points, to which the coils must add flux to equal the LCFS value. We then minimize the difference between this desired value and the flux created by the optimized coils:

\begin{equation}
    \Phi_\textrm{div} = \sum_j w_j^\textrm{div} \left[\psi_\textrm{LCFS}^\textrm{opt} - \psi_j^\textrm{plasma} - \sum_iG_{ji}I_i\right]^2
\end{equation}

where $w_j$ is another inverse square weighting, typically found through experimentation.

A similar optimization term can be created to in order to create flux expansion along the separatrix as it approaches the strike point, this is done by a similar method but using the magnetic flux density:

\begin{equation}
    \Phi_\textrm{div,flux} = \sum_j w_j^\textrm{div,flux} \left|\underline{B}_j^\textrm{plasma} + \sum_i\left(G_{ji}^z\underline{\hat{z}} + G_{ji}^r\underline{\hat{r}}\right)\underline{G}_{ji}I_i\right|^2
\end{equation}

which requires the flux density from the plasma at the chosen points, $\underline{B}_j^\textrm{plasma}$, and the Green's function for the flux density from the coils,  $\underline{G}_{ji}$. This is not broken down into components as we wish to minimize the component perpendicular to the isoflux separatrix, which is the bare field by definition.

\section{Cost-function Calculation Methods}
Here we will briefly discuss the general nature of the above constraint and cost-function terms and using them in an analytic optimization. Firstly, they are all chosen to be quadratic or quartic with current as this allows the terms and their derivatives to be evaluated in a very straightforward manner using matrices and vertices, typically combining lots of similar terms.

Additionally, optimizing for the L2 norm of each physical term will ensure that the higher load cases are more penalized so that the total load isn't reduced at the cost of increased loads on the most stressed coils, but instead more evenly distributed, even if the total is higher.

In the above terms, we normalize by dividing by the square of a desired value. This creates dimensionless terms that can then be added together, but also, as there are many separate terms, we can start with the desired values as reasonable initial guesses which can then be increased or decreased based on feedback from the optimization results or from more detailed engineering design and analysis.

In order to calculate the magnetic terms above, we use an integral method that relies on the cylindrical symmetry of the system and uses elliptic integrals to calculate the flux and fields created by infinitely thin current filaments \cite{Garrett1963EllipticIntegrals}. Multiple filaments are then used to integrate over the coil cross-section using Gaussian quadrature. For the forces that require integrating over the cross-section of the coil, we use the flux-field relationship to reduce the integration to two 1D integrals along the edges of the coil.

In order to produce a computationally tractable problem, the simulated annealing method allows the various optimization matrices and vectors to be updated rather than recalculated in full.

\section{Coupling to Simulated Annealing}

Simulated annealing works by applying random perturbations to a system and then selectively keeping or dismissing the perturbed system depending on the change to the performance \cite{Press2007NumericalRecipes}. Perturbations that caused improvements are always kept, costly perturbations are randomly kept with a probability exponentially dependent on the size of the performance decrease relative to a pseudo-temperature. If a perturbation is not kept the system is returned to it's previous state and another perturbation is performed. By carrying out a great many perturbations with a steadily decreasing temperature the system performance should increase over time. This method allows for global optimization as initially a system can climb out of any local minima it may have started in and then search the entire parameter landscape whilst preferentially selecting higher performing areas.

To allow the fast update method, every iteration in the simulated annealing method randomly choses a single coil, then a randomly chosen parameter, $R$, $Z$, $\Delta R$ or $\Delta Z$, is shifted by a small random amount with even distribution and maximum variation of $\pm2.5$ mm. Note that spatially constrained coils such as the central solenoid are not included in this step but are included in the cost-function and current optimization.

After randomly altering a coil, any simple constraints are checked such as coil-coil overlap or coil-spatial constraint as these are fast and can quickly allow a non-feasible change to be rejected without any further analysis. Additional limits such as maximum or minimum coil size or position are enforced by mirroring the over or under shoot eg if a coil has it's $R$ position increased by a random amount outside the feasible range, $R_\textrm{max}$, then we use $R \rightarrow 2R_{max} - R$. This is preferable to simply stating $R = R_\textrm{max}$, as this would produce a redundant trial if $R$ was already at the maximum and will also increase and not decrease the movement of any coil property that has hit a spatial limit, in that any change will help move it away from that limit.

If all hard spatial constraints are passed, then the plasma constraint and optimization matrices, vectors and constants are updated for the chosen coil and all it's interactions. These values are then passed to an interior point constrained optimization that attempts to minimize the following barrier Lagrangian \cite{InteriorPointMethods}:

\begin{equation}
    \mathcal{L} = \Phi(I) - \mu\sum_i \log[C_i(I)] 
\end{equation}

where $\Phi$ is the sum of all cost-function terms, either in time with additional weightings depending on timestep between static equilibria or just summed if over multiple flat-top equilibria, and $\mu$ is an additional weighting function for the logarithmic barrier term which ensures the constraint terms cannot become negative or zero and should become negligible when the constraints are satisfied. Though this is a standard method, the implementation is written to make use of simplifications and repetition in the matrix-vector calculations, rather than relying on an external black-box method.

This process also produces a set of Lagrangian multipliers, $\lambda_i = \mu/C_i$, which are kept and used for the next iteration. Also, helpfully, we can use the relationship at optimality: $\lambda_i  = d\Phi/dC_i$ which gives an indication for the robustness of the solution, ie how the performance of the coils will change if the constraint function changes. These can help show that any optimal solution we obtain is not too sensitive to change in the plasma scenario. As we normalize our constraint function so that 0.25\% is the required tolerance, we can use the cost-function for each equilibrium: $\Phi_i / \lambda_i$ as a proportional measure of how much the cost-function will increase proportional to a change in constraint as a large change in the constraint is likely to be of order of inverse tolerance, so 4000. 

This then produces the optimum currents and cost-function for the randomly perturbed coils. To this current based cost-function, additional terms can be added for any broken spatial constraints and the simple volume terms.

The simulated annealing algorithm will then allow this random change to the coil using the standard selection method:

\begin{equation}
    R > \exp[-\Delta\Phi/T]
\end{equation}

where $R$ is a random number uniformly chosen from 0 to 1, $\Delta \Phi$ is the change in cost-function caused by the random change, and $T$ is a pseudo-temperature that decreases throughout the optimization. If the change does not pass this selection method then it is reverted and the coils are returned to their previous state, ready for another change. In practice, we keep the original coils and associated data and create a copy that has the random change applied to it, then if the change is accepted, we update the original copy. We also keep a copy of the optimum coils at all points which is only updated if the cost-function has decreased below the previous minimum.

As we can see from the selection method, if $\Delta\Phi$ is negative, ie the random change has helped to reduce the cost-function and improved performance, then the selection will always pass. However, if $\Delta\Phi$ is positive, and the coils are now performing worse, then the selection is not guaranteed and exponentially dependent on how much the change degraded the coils and what the pseudo-temperature is currently set at.

For our annealing algorithm, the pseudo-temperature is linearly decreased for each iteration, from experiment this was found to be the simplest 'cooling schedule' and by tracking 'steps' made up of 50 individual trials, we can get a good idea of how many trials are accepted, how many don't meet the spatial constraints and how many trials improve on the previous cost function. A heuristic method was used to set the initial temperature which was to get $>75\%$ of the first few steps to be accepted to allow the initial coils to climb out of any local minima. The linear decrease in temperature was always set so that very last step was one iteration from 0, so that at the end of the optimization only negative $\Delta\Phi$ were likely to be selected. lastly, we set the length of the optimization with another heuristic, whereby the percentage of accepted changes in each step of 50 iterations should not show a 'quench' where they suddenly drop off exponentially. If this happened the optimization was re-run with double the number of steps. We found that 1,000 steps would allow a reasonable optimization without too sharp of a quench, but 30,000 steps (1.5 million iterations) was required for a more confident optimization that typically improved the performance by a further ~10\%. Exponential cooling was tried, though it was found that for the same initial temperature and length of optimization, cooling happened too fast near the start and coils would get stuck in coupled local optima.

Large numbers of iterations are required for this optimization method due it's stochastic nature, however, this allows us to vary a large number of degrees of freedom and search an incredibly large parameter space, 40 dimensions (20 spatial and 20 currents) for the example given below, whilst simultaneously optimizing for $\sim$50 to 100 equilibria. Written in MATLAB, running on a laptop Intel® Core™ i7-12800H CPU (14 cores, 2.4 GHz), each coil and current update would take less than 300ns per equilibrium.  Therefore, a typical short optimization of 1,000 steps to test the effects of optimization weights or temperatures etc would take less than 15 minutes and a more confident optimization, 30,000 steps, to get final results would take around 7 hours. 

\section{Results}
The following discussion of results is not an exhaustive examination of all the individual terms, instead we show a few representative case studies of applying the optimization methodology to an iteration of the ST-E1 device.

\subsection{Simultaneous Equilibria Optimization}
The first case study, used early in the pre-concept design stages was termed the 'smudge' which was a collection of steady-state equilibria with varying plasma parameters. By varying quantities such as elongation, triangularity, plasma current, and pedestal size, we cover a large potential solution space for the final plasma design. This decouples magnet design and physics investigations, and allows us to design a magnet cage which is able to handle a wide range of plasma scenarios, giving us time to investigate and down-select an optimal operating point. 

For this spread of simultaneous flat top scenarios, we carried out the optimization keeping the LCFS flux exactly the same as the FBE value so as to directly compare the effect of the optimization on the coil performances. Shown in table \ref{tb:singleCase} is the current, max field and average hoop stress in each coil, first for the currents in the initial FBE and then after the coils and currents have been optimized for one of the scenarios shown in fig \ref{fig:}. We can see that the currents are drastically reduced for PF1 which was previously in excess of 100MA-turns, similarly the solenoid now requires 1/3 of its previous current. Conversely, PF3, which previously was using very little current has increased due to the sharing of the loads across the coils. We see similar effects for the maximum magnetic field on the coil perimeter, the effect has increased as not only have the larger currents decreased but also the coils have increased in size, especially for PF5. These two effects are then compounded for the hoop stress due to the Lorentz force so we can see massive reduction in all coils, this is critical as given that the yield strength of copper, a key component in the coils, is around 300MPa \cite{CopperProperties} and after optimization we have reached this for all coils. In table \ref{tb:allCases} we show the same result, taking a root mean square of the coil performances over all 71 cases used in the optimization. We see similar results again though with the more extreme cases averaged out, though again we see the destructively high hoop stresses reduced to very managable values. Interestingly, as there was no flux swing, the solenoid was the same on average.

\begin{table}[h!]
\label{tb:singleCase}
\centering
\setlength{\tabcolsep}{1pt} 
\footnotesize
\caption{Comparison of Initial and Optimized Coil Parameters for a single scenario from fig \ref{fig:}}
\begin{tabular}{|c|c|c|c|c|c|c|c|c|c|}
\hline
\makecell{Coil\\Name} &
\makecell{Initial\\Current\\(MA·turns)} &
\makecell{Optimized\\Current\\(MA·turns)} &
\makecell{Factor\\Reduction} &
\makecell{Initial\\Max Field\\(T)} &
\makecell{Optimized\\Max Field\\(T)} &
\makecell{Factor\\Reduction} &
\makecell{Initial\\Max Hoop\\Stress (MPa)} &
\makecell{Optimized\\Max Hoop\\Stress (MPa)} &
\makecell{Factor\\Reduction} \\
\hline
PF1       & 143   & 4.36   & 32.7   & 79.5  & 7.13  & 11.2   & 2740  & 30.2   & 90.6  \\
PF2       & -38.1 & 35.3   & 1.08   & 34.1  & 12.4  & 2.75   & 1470  & 6.55   & 225   \\
PF3       & -8.82 & -28.8  & 0.306  & 5.33  & 10.2  & 0.525  & -104  & 30.0   & 3.45  \\
PF4       & 42.1  & 34.4   & 1.22   & 20.7  & 10.1  & 2.04   & -1360 & -21.3  & 64.2  \\
PF5       & -43.4 & -31.9  & 1.36   & 22.0  & 7.28  & 3.03   & 2260  & 110    & 20.7  \\
Solenoid  & 55.4  & 18.8   & 2.95   & 4.03  & 2.50  & 1.61   & 3.10  & 6.84   & 0.448 \\
\hline
\end{tabular}
\end{table}

\begin{table}[h!]
\label{tb:allCases}
\centering
\setlength{\tabcolsep}{1pt} 
\footnotesize
\caption{Comparison of Initial and Optimized Coil Parameters using over all scenarios shown in fig \ref{fig:}}
\begin{tabular}{|c|c|c|c|c|c|c|c|c|c|}
\hline
\makecell{Coil\\Name} &
\makecell{Initial\\Current\\(MA·turns)} &
\makecell{Optimized\\Current\\(MA·turns)} &
\makecell{Factor\\Reduction} &
\makecell{Initial\\Max Field\\(T)} &
\makecell{Optimized\\Max Field\\(T)} &
\makecell{Factor\\Reduction} &
\makecell{Initial\\Max Hoop\\Stress\\(MPa)} &
\makecell{Optimized\\Max Hoop\\Stress\\(MPa)} &
\makecell{Factor\\Reduction} \\
\hline
PF1       & 86.2  & 4.87   & 17.7   & 48.3  & 6.21  & 7.78   & 1670  & 30.6  & 54.7  \\
PF2       & 29.6  & 25.2   & 1.18   & 22.9  & 9.15  & 2.50   & 882   & 33.0  & 26.7  \\
PF3       & 16.4  & 17.9   & 0.917  & 9.18  & 6.73  & 1.36   & 507   & 75.0  & 6.76  \\
PF4       & 34.8  & 17.1   & 2.03   & 16.7  & 5.66  & 2.95   & 904   & 62.1  & 14.5  \\
PF5       & 24.7  & 19.9   & 1.24   & 12.7  & 4.37  & 2.90   & 1158  & 50.8  & 22.8  \\
Solenoid  & 21.1  & 13.3   & 1.59   & 4.03  & 2.50  & 1.33   & 3.10  & 6.84  & 0.448 \\
\hline
\end{tabular}
\end{table}

\subsection{Ramp-up optimization}
The second case study we wish to show is the optimization of a ramp-up scenario, shown using the flux swing in figure \ref{fig:fluxVals}. As explained above, here we take a set of static FBE equilibria and we optimize a set of coils with time varying currents that will not only hold the plasma but also apply a flux swing while removing the FBE flux increase. We have shown above the coils in the ST-E1 machine are capable of holding a plasma but this flux swing will put much greater stresses on the coils.

\begin{figure} [ht!]
\centering
\includegraphics[width=0.5\linewidth]{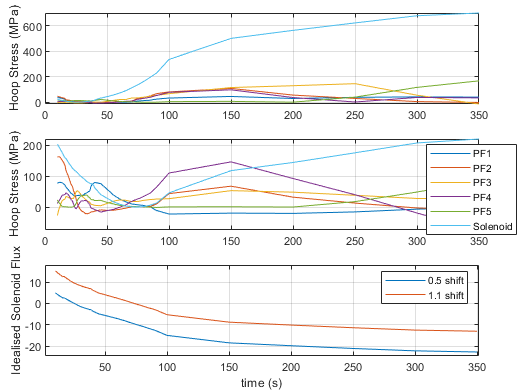}
\caption{\label{fig:ramp} (Top) hoop stresses in the coils with a 30Vs LCFS flux swing centred on zero, very large central solenoid stresses can be seen at the end of the swing due to confinement of a large plasma at the end of the ramp. (Middle) hoop stresses with a LCFS flux swing shifted by (+33Vs) creating greatly reduced stresses. (Bottom) Idealised solenoid flux, ie the total flux through the centre of the solenoid, for the symmetric and shifted flux swing. While the flux swing on the plasma may be symmetric, we can see that the optimal solenoid swing is not and vice versa.}
\end{figure}

We first begin by optimizing assuming a symmetric positive-negative flux swing and so we take the flux values from METIS and shift them by exactly a half of the total Volt-seconds. As shown in figure \ref{fig:ramp} (top), when we optimize this system, we find that the solenoid suffers very large stresses at the end of the ramp. To reduce this, we instead shift the flux swing on the LCFS by 1.1 times the flux swing, therefore the LCFS never becomes negative, in this case as shown in figure \ref{fig:ramp} (middle) the hoop stresses on the solenoid are greatly reduced and become similar to those of the PF coils. If we look the flux passing through the bore of the solenoid, using the idealised solenoid, figure \ref{fig:ramp} (bottom) shows that originally setting the LCFS flux to follow a symmetric positive-negative swing corresponds to the solenoid undergoing a very asymmetric swing and when shifting the LCFS swing to reduce the strain on the solenoid, we then retrieve a symmetric positive-negative swing on the solenoid. This is inline with the typical bi-polar swing design of most tokamak solenoids, eg \cite{ARC_overview_2015}, however here we have taken into account the PF-CS-plasma coupling to ensure that the LCFS flux swing is still following the desired flux swing given by all field sources and minimizing the issues with all coil interactions.

\section{Conclusion}
Here we have outline an optimization method to minimize coil engineering issues whilst working collaboratively with a plasma and divertor scenario development. The method allows for large numbers of coil design parameters to be found whilst ensuring that hard constraints like plasma confinement and spatial limitations are met. This likely reduces the loads on the coils and converts previously unacceptable FBE results into feasible solutions, without at all altering the plasma performance. The output of the optimization is also deliberately set up to allow for feedback on what is possible for the coils as well as what scenarios are more taxing or entirely unfeasible and what alterations may be possible. This allows plasma scenario development to continue with a large amount of feedback, rather than just a go/no-go flag. What greatly helps is the ability to consider multiple equilibria simultaneously so parameter scans and heat maps can be created. Additionally, further engineering such as structure, cryogenics and power supply can also use the output of the optimization for further detailed design work and feedback on further optimization via alteration of optimization weights or adding in additional cost-function terms.

\section{Acknowledgements}
The pre-concept design study of ST-E1 and targeted improvements to the SCOPE optimization tool were undertaken with support from the U.S. Department of Energy through the Milestone-Based Fusion Development Program.

\bibliographystyle{ieeetr}
\bibliography{sample}

\begin{thebibliography}{10}

\bibitem{STE1_overview_paper_PLACEHOLDER}
J.~Willis {\em et~al.}, ``St-e1 pre-concept overview,'' {\em To be submitted}, 2026.
\newblock Placeholder for ST-E1 overview paper.

\bibitem{STE1_flat_top_paper_PLACEHOLDERR}
S.~A.~M. McNamara {\em et~al.}, ``St-e1 flat-top overview,'' {\em To be submitted}, 2026.
\newblock Placeholder for ST-E1 flat-top paper.

\bibitem{STE1_ramp_up_PLACEHOLDER}
X.~Zhang, N.~A. Lopez, M.~Borscz, M.~Scarpari, J.~Kang, S.~A.~M. McNamara, Y.~Takase, C.~Marsden, E.~N.~J. Maartensson, M.~Ono, {\em et~al.}, ``Investigation of time-dependent plasma ramp-up scenarios for st-e1,'' {\em To be submitted}, 2026.
\newblock Placeholder for ST-E1 ramp-up paper.

\bibitem{Metis_overview_2018}
J.~Artaud, F.~Imbeaux, J.~Garcia, G.~Giruzzi, T.~Aniel, V.~Basiuk, A.~Bécoulet, C.~Bourdelle, Y.~Buravand, J.~Decker, R.~Dumont, L.~Eriksson, X.~Garbet, R.~Guirlet, G.~Hoang, P.~Huynh, E.~Joffrin, X.~Litaudon, P.~Maget, D.~Moreau, R.~Nouailletas, B.~Pégourié, Y.~Peysson, M.~Schneider, and J.~Urban, ``Metis: a fast integrated tokamak modelling tool for scenario design,'' {\em Nuclear Fusion}, vol.~58, p.~105001, aug 2018.

\bibitem{FreeGS_documentation_2025}
B.~Dudson, {\em FreeGS: Free-boundary Grad--Shafranov Solver}.
\newblock University of York, 2022.
\newblock Accessed: 2025-10-08.

\bibitem{Metis_FEEQS_coupling_2021}
V.~Ostuni, J.~F. Artaud, G.~Giruzzi, E.~Joffrin, H.~Heumann, and the JT-60SA Modelling~Team, ``Tokamak discharge simulation coupling free-boundary equilibrium and plasma model with application to jt-60sa,'' {\em Nuclear Fusion}, vol.~61, p.~026021, 2021.
\newblock HAL Id: cea-03287071.

\bibitem{M_Windridge_TE_and_HTS_spherical_tokamaks_overview_2019}
M.~Windridge, ``Smaller and quicker with spherical tokamaks and high-temperature superconductors,'' {\em Philosophical Transactions of the Royal Society A: Mathematical, Physical and Engineering Sciences}, vol.~377, no.~2141, p.~20170438, 2019.

\bibitem{STEP_special_issue_TheMagneticCage}
E.~Nasr, S.~C. Wimbush, P.~Noonan, P.~Harris, R.~Gowland, and A.~Petrov, ``The magnetic cage,'' {\em Philosophical Transactions of the Royal Society A: Mathematical, Physical and Engineering Sciences}, vol.~382, p.~20230407, 2024.
\newblock One contribution of 16 to the theme issue ‘Delivering Fusion Energy – The Spherical Tokamak for Energy Production (STEP)’.

\bibitem{ARC_overview_2015}
B.~Sorbom, J.~Ball, T.~Palmer, F.~Mangiarotti, J.~Sierchio, P.~Bonoli, C.~Kasten, D.~Sutherland, H.~Barnard, C.~Haakonsen, J.~Goh, C.~Sung, and D.~Whyte, ``Arc: A compact, high-field, fusion nuclear science facility and demonstration power plant with demountable magnets,'' {\em Fusion Engineering and Design}, vol.~100, pp.~378--405, 2015.

\bibitem{HH70_overview_2025}
Z.~Yang, Y.~Cao, B.~Chen, G.~Chen, W.~Chen, G.~Dong, Y.~Du, H.~Guo, Y.~Guo, Y.~Hua, Y.~Huang, Y.~Li, Z.~Li, J.~Liu, J.~Ma, Z.~Pan, H.~Qiao, Y.~Wang, Q.~Wei, H.~Yang, Y.~Ye, S.~Zhai, Y.~Zhao, C.~Zhang, K.~Zhang, Y.~Zhang, Z.~Zhang, X.~Zhou, K.~Zhu, and Y.~Zhu, ``Design, commissioning, and first operation of the high-temperature superconducting tokamak honghuang70,'' {\em Fusion Engineering and Design}, vol.~220, p.~115341, 2025.

\bibitem{sparc_overview_2020}
A.~J. Creely, M.~J. Greenwald, S.~B. Ballinger, D.~Brunner, J.~Canik, J.~Doody, T.~Fülöp, D.~T. Garnier, R.~Granetz, T.~K. Gray, and et~al., ``Overview of the sparc tokamak,'' {\em Journal of Plasma Physics}, vol.~86, no.~5, p.~865860502, 2020.

\bibitem{sparc_TF_program_2024}
Z.~S. Hartwig, R.~F. Vieira, D.~Dunn, T.~Golfinopoulos, B.~LaBombard, C.~J. Lammi, P.~C. Michael, S.~Agabian, D.~Arsenault, R.~Barnett, M.~Barry, L.~Bartoszek, W.~K. Beck, D.~Bellofatto, D.~Brunner, W.~Burke, J.~Burrows, W.~Byford, C.~Cauley, S.~Chamberlain, D.~Chavarria, J.~Cheng, J.~Chicarello, V.~Diep, E.~Dombrowski, J.~Doody, R.~Doos, B.~Eberlin, J.~Estrada, V.~Fry, M.~Fulton, S.~Garberg, R.~Granetz, A.~Greenberg, M.~Greenwald, S.~Heller, A.~E. Hubbard, E.~Ihloff, J.~H. Irby, M.~Iverson, P.~Jardin, D.~Korsun, S.~Kuznetsov, S.~Lane-Walsh, R.~Landry, R.~Lations, R.~Leccacorvi, M.~Levine, G.~Mackay, K.~Metcalfe, K.~Moazeni, J.~Mota, T.~Mouratidis, R.~Mumgaard, J.~Muncks, R.~A. Murray, D.~Nash, B.~Nottingham, C.~O'Shea, A.~T. Pfeiffer, S.~Z. Pierson, C.~Purdy, A.~Radovinsky, D.~K. Ravikumar, V.~Reyes, N.~Riva, R.~Rosati, M.~Rowell, E.~E. Salazar, F.~Santoro, A.~Sattarov, W.~Saunders, P.~Schweiger, S.~Schweiger, M.~Shepard, S.~Shiraiwa, M.~Silveira, F.~Snowman, B.~N. Sorbom, P.~Stahle, K.~Stevens,
  J.~Stillerman, D.~Tammana, T.~L. Toland, D.~Tracey, R.~Turcotte, K.~Uppalapati, M.~Vernacchia, C.~Vidal, E.~Voirin, A.~Warner, A.~Watterson, D.~G. Whyte, S.~Wilcox, M.~Wolf, B.~Wood, L.~Zhou, and A.~Zhukovsky, ``The sparc toroidal field model coil program,'' {\em IEEE Transactions on Applied Superconductivity}, vol.~34, no.~2, pp.~1--16, 2024.

\bibitem{SuperconComparison}
P.~J. Lee, ``A comparison of superconductor critical currents,'' July 2024.
\newblock Accessed: 21 Oct 2025.

\bibitem{EAST_PF_optimisation_genetic_algorithms_2006}
Z.~L. An, X.~P. Liu, B.~Wu, and X.~J. Zha, ``Optimization of positions and currents of tokamak poloidal field coils using genetic algorithms,'' {\em Fusion Science and Technology}, vol.~50, no.~4, pp.~561--568, 2006.

\bibitem{PF_optimisation_using_BLUEPRINT_2020}
M.~Coleman and S.~McIntosh, ``The design and optimisation of tokamak poloidal field systems in the blueprint framework,'' {\em Fusion Engineering and Design}, vol.~154, p.~111544, 2020.

\bibitem{Nilima2024_FusionEngDes_STEP_Bluemira_PFcoil_Optimisation}
A.~Nilima, M.~Bluteau, J.~Cook, O.~Funk, G.~Graham, J.~Matthews, S.~I. Muldrew, A.~J. Pearce, and D.~Vaccaro, ``Optimisation of step poloidal field coils with superconducting coil constraints in step-bluemira power plant design framework,'' {\em Fusion Engineering and Design}, 2024.
\newblock Available at ScienceDirect: \url{https://www.elsevier.com/locate/fusengdes}.

\bibitem{bayesian_optimisation_of_PF_coils_STEP_2025}
T.~Nunn, K.~Pentland, V.~Gopakumar, and J.~Buchanan, ``Bayesian optimization of poloidal field coil positions in tokamaks,'' {\em Physics of Plasmas}, vol.~32, no.~7, p.~072507, 2025.

\bibitem{Bardsley2024_PPCF_Decoupled_Magnetic_Control_ST_Divertors}
O.~P. Bardsley, J.~L. Baker, and C.~Vincent, ``Decoupled magnetic control of spherical tokamak divertors via vacuum harmonic constraints,'' {\em Plasma Physics and Controlled Fusion}, vol.~66, no.~5, p.~055006, 2024.
\newblock © 2024 Crown copyright, UKAEA. Open access.

\bibitem{STE1_integration_paper_PLACEHOLDER}
E.~N. J.~M. et~al, ``Integrated physics and magnet development of st-e1,'' {\em To be submitted}, 2026.
\newblock Placeholder for ST-E1 integrated magnets and plasma paper.

\bibitem{ErnshawTheorem}
J.~Jeans, {\em General analytical theorems, in Mathemati- cal Theory of Electricity and Magnetism, p. 156{184, 5th ed}}.
\newblock Physical Sciences (Cambridge University Press), 2009.

\bibitem{Garrett1963EllipticIntegrals}
M.~W. Garrett, ``Calculations of fields, forces, and mutual inductances of current systems by elliptic integrals,'' {\em Journal of Applied Physics}, vol.~34, no.~9, pp.~2567--2573, 1963.

\bibitem{Press2007NumericalRecipes}
W.~H. Press, S.~A. Teukolsky, W.~T. Vetterling, and B.~P. Flannery, {\em Numerical Recipes: The Art of Scientific Computing}.
\newblock Cambridge, UK: Cambridge University Press, 3~ed., 2007.

\bibitem{InteriorPointMethods}
S.~J. Wright, {\em Primal‑Dual Interior‑Point Methods}.
\newblock SIAM, 1997.

\bibitem{CopperProperties}
AALCO, ``Copper and copper alloys.''
\newblock Accessed: 21 Oct 2025.

\end{thebibliography}

\end{document}